\documentclass[12pt,prb,aps,preprint]{revtex4}

\begin{document}

\title{An elliptic  property of  parabolic trajectories}

\author{J. L. Fern\'andez-Chapou,  A. L. Salas-Brito, C. A. Vargas}
\affiliation{Departamento de Ciencias B\'asicas, Universidad
Aut\'onoma Metropolitana-Azcapotzalco, Apartado
Postal 21-267, Coyoac\'an 04000 D.\ F., M\'exico}
\email{jlfc@correo.azc.uam.mx,asb@correo.azc.uam.mx,cvargas@correo.azc.uam.mx}

\begin{abstract}
The curve joining the points of maximum height in the parabolas of ideal projectile motion is shown  to be an ellipse. Some features of the motion are illustrated with the help of such ellipse.
\end{abstract}

\maketitle

 In the last  years several contributions  have analysed  the old but ever interesting problem  of the properties of projectile motion. \cite{bose,bajc,french,price,bace,butikov} In this note we want to pinpoint that  an ellipse is also related to the  parabolas traced by projectiles launched in a constant  field. The ellipse appears as the curve connecting the points of maximum height in the parabolas. 

To  obtain this little known fact,  recall that the coordinates of projectiles  launched from the origin in the $x$--$y$ plane with  the same initial speed $v_0$ but different launching angles  are

\begin{equation}\label{path}
x=v_0 t\cos\alpha , \qquad y=v_0 t \sin\alpha -\frac{1} {2} gt^2,
\end{equation} 
 
\noindent where $\alpha$ is the launching angle. From these equations it is simple to obtain the time for reaching the maximum height as $t_m= v_0\sin\alpha/g$. Therefore, the coordinates of the projectile at $t_m$  are

\begin{equation}\label{tm}
x_m= \frac {v_0^2}{g} \cos\alpha \sin\alpha, \qquad y_m= \frac{1}{2} \frac{v_0^2}{g} \sin^2\alpha,
\end{equation}

\noindent where $y_m$ is the maximum height attained and $x_m$ the corresponding horizontal coordinate. It should be noted that the trajectories described by Eq.  (\ref{path}) are symmetric respect to the point $(x_m, y_m)$ and so  the  range of a projectile launched at an angle $\alpha$ is $R=2x_m=2\cos \alpha \sin\alpha \,v_0^2/g$.
With the help of the trigonometric identities $2\cos\alpha\sin\alpha=\sin2\alpha$ and $2\sin^2\alpha=1-\cos2\alpha$, Eqs.\ (\ref{tm}) can be casted in the form

\begin{equation}\label{h}
x_m= \frac {v_0^2}{2g}  \sin2\alpha, \qquad y_m=  \frac{v_0^2}{4g} (1-\cos 2\alpha).
\end{equation}

\noindent Eliminating the angle $\alpha$ from (\ref{h}) we get the loci of points of maximum height

\begin{equation}\label{ellipse}
\frac{x_m^2}{a^2}+\frac{(y_m-b)^2}{b^2}=1,
\end{equation}

\noindent where we have defined $a=v_0^2/2g$, and  $b=v_0^2/4g$, {\it i. e.} $a=2b$. Eq.\ (\ref{ellipse}) represents an ellipse centered at point $(0, b)$ with minor and major axes given,  respectively, by  $2b$ and $2a$.  The eccentricity is  $e=\sqrt{3}/2 $  and does not depend on any details of the motion.  See Fig.\ 1 for a graph of the ellipse and of some of the projectile trajectories that generate it.  

 Discussing different ways of obtaining a result  may  enhance student retention and understanding.   To obtain (\ref{ellipse}) in another way,  we can  start from the gedanken experiment of simultaneously firing from the origin  many projectiles with the same initial speed $v_0$---{\it e. g.} the case of an exploding firecracker.\cite{butikov}    At any later time, $t$, the projectiles are on the curve

\begin{equation}\label{circle}
x^2 + \left(y+{1\over 2} g t^2\right)^2=v_0^2t^2.
\end{equation}

\noindent That is, the projectiles are located on  an expanding and freely falling circle. To see it,  take the point of view of a free-falling frame  where gravity vanishes \cite{taylor}. In this frame  the projectiles are  on a expanding circle of radius $v_0 t$---as can be  derived from Eqs.\ (\ref{path}) with $g=0$. But  in the frame where gravity is acting,  the center of mass is falling at a constant acceleration $g\neq 0$, hence  it is located  at (0, $-gt^2/2$) and Eq.\ (\ref{circle}) follows. When we substitute the time  $t_m$ in this equation we  obtain that the points of maximum height, ($x_m$, $y_m$), satisfy

\begin{equation}\label{ellipse2}
\quad x_m^2+ \left(y_m+{1\over 2} {v_0^2\over g}\sin^2\alpha\right)^2={v_0^4\over g^2}\sin^2\alpha.
\end{equation}

\noindent  To eliminate  $\sin^2\alpha$ from (\ref{ellipse2}) we  employ  Eq.\ (3) and obtain Eq.\ (\ref{ellipse}) again. In fact, we can do an identical calculation making not reference whatsoever to particles on a circle, since the equation of the path of a {\sl single} projectile in the  falling frame   is $x'^2+y'^2=v_0^2t^2 $ ---the primed variables being coordinates in such frame.   

Our ellipse  also affords yet another solution to the problem of maximizing the range in projectile motion since the right vertex of it (see Fig.\ 1)  corresponds precisely to such a case.

Can the ellipse (\ref{ellipse}) be characterized in any other way?  It can be shown to be also the locus of points where  the radial ($\dot r$) component of the velocity vanishes.  Stablishing this property can be a simple exercise in intermediate mechanics. It can also be proved that as long as the downward  branch ({\it i.\ e.\ } the one in which $\dot y<0$) of a projectile path is within the ellipse then the projectile is necessarily approaching the origin otherwise it is getting away from it.

\begin{acknowledgments}
ALSB has been partially supported by a PAPIIT-UNAM grant 108302. Helpful comments from D Crida, H Kranken,  and K Bielii are  gratefully acknowledged.  Gabriela B\'aez is  acknowledged for spotting a gross  misprint in the typescript. This work is  for  Mileva Sof\'{\i}a.
\end{acknowledgments}

\newpage
\begin{figure}[h]
\caption{The ellipse of maximum heights associated with a family of parabolic trajectories  launched with  the same initial speed (thick black curve). The center of the ellipse is at $(0, b)$ and its eccentricity is $\sqrt{3}/2$.  The ten lighter curves are some of the projectile paths giving birth to it. The gray curve is  the parabolic envelope of the family of projectile trajectories. The numbers shown on the right of the $y$-axis correspond to the launching angle of the nearest parabola. The symmetric parabolas on the left correspond to  projectiles launched with the same speed but complementary angles. This plot makes apparent that the maximum range is reached for $\alpha=45^\circ$, {\it i. e.} it corresponds to the right vertex of the ellipse, and is   $R_m= 2a$.}
\end{figure}
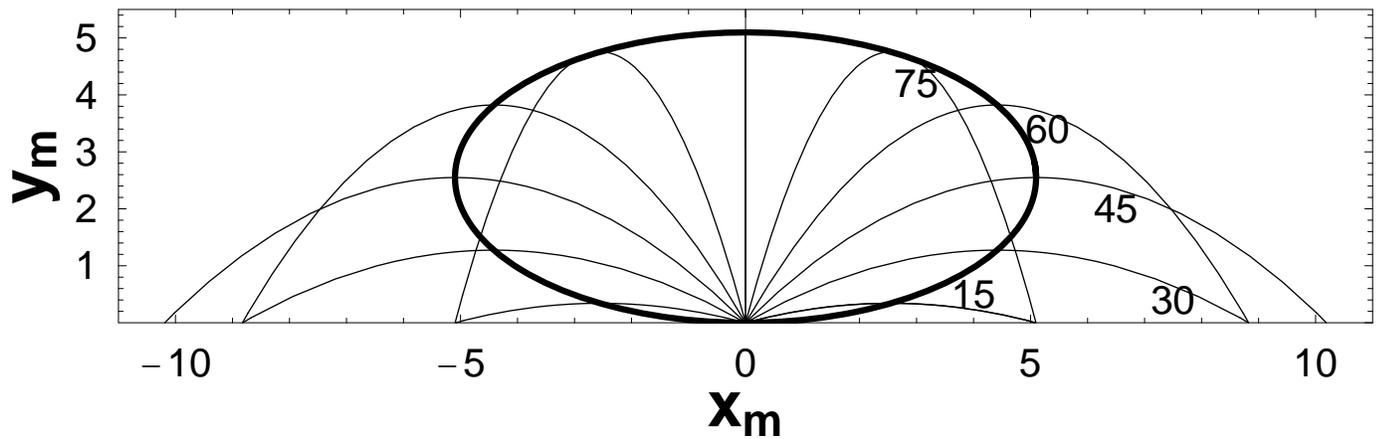
\end{document}